# Synthetic Turing protocells: vesicle self-reproduction through symmetry-breaking instabilities

Javier Maciá[1] and Ricard V. Solé[1,2]

[1]ICREA-Complex Systems Lab, Universitat Pompeu Fabra (GRIB), Dr Aiguader 80, 08003 Barcelona, Spain
[2]Santa Fe Institute, 1399 Hyde Park Road, Santa Fe NM 87501, USA

The reproduction of a living cell requires a repeatable set of chemical events to be properly coordinated. Such events define a replication cycle, coupling the growth and shape change of the cell membrane with internal metabolic reactions. Although the logic of such process is determined by potentially simple physico-chemical laws, the modeling of a full, self-maintained cell cycle is not trivial. Here we present a novel approach to the problem which makes use of so called symmetry breaking instabilities as the engine of cell growth and division. It is shown that the process occurs as a consequence of the breaking of spatial symmetry and provides a reliable mechanism of vesicle growth and reproduction. Our model opens the possibility of a synthetic protocell lacking information but displaying self-reproduction under a very simple set of chemical reactions.

## I. INTRODUCTION

In 1952 Alan Turing published a very influential paper in Philosophical Transactions of the Royal Society, entitled *The Chemical Basis of Morphogenesis*. In that article Turing proposed a possible solution to the problem of how developing systems can become heterogeneous, spatially organized entities starting from an initially homogeneous state (Turing 1952, Meinhardt 1982, Murray 1989, Lengyel and Epstein, 1992). Turing showed that an appropriate compromise between local reactions and long-range communication through diffusion could generate macroscopic spatial structures. The interplay of both components was described in terms of a system of partial differential equations, so called *reaction-diffusion* (RD) equations, namely a set

$$\frac{\partial C_1}{\partial t} = \Phi_1(C_1, C_2) + D_1 \nabla^2 C_1 \qquad (1)$$

$$\frac{\partial C_2}{\partial t} = \Phi_2(C_1, C_2) + D_2 \nabla^2 C_2 \qquad (2)$$

Here, $C_1$ and $C_2$ indicate the concentrations of the two morphogens and their specific molecular interaction are described by the reaction terms $\Phi_i(C_1, C_2)$ with $i = 1, 2$. These reactions could be, for example, activations, inhibitions or autocatalysis and degradation. The concentrations are spatially and time-dependent functions i. e. $C_1 = C_1(r, t)$ and $C_2 = C_2(r, t)$. Here $r$ indicates the coordinates of a point $r \in \Gamma$ where $\Gamma$ is the spatial domain where reactions occur.

The last terms in the right hand side of both equations stand for diffusion over space: here $D_1$ and $D_2$ are the corresponding diffusion rates, indicating how fast each molecule diffuses through space. If $\Gamma$ is a two-dimensional area, we would have $r = (x, y)$ and the diffusion term reads:

$$\nabla^2 C_i = \frac{\partial^2 C_i}{\partial x^2} + \frac{\partial^2 C_i}{\partial y^2} \qquad (3)$$

which can be properly discretized using standard numerical methods. The key idea of Turing instabilities is that small initial fluctuations can be amplified through the reaction terms (typically nonlinear) and their effects propagate through space thanks to diffusion. These patterns are generated from an initially almost homogeneous distribution of morphogens. Specifically, we use as a reference state the equilibrium concentrations $C_1^*, C_2^*$ obtained from the condition $dC_1/dt = dC_2/dt = 0$. It is not difficult to see that if the system starts exactly from this homogeneous state, it will remain there forever. Now consider a very small perturbation of such homogeneous state, where the initial concentration is now: $C_i(r, 0) = C_i^* + \xi(r, 0)$ with $\xi(r, 0)$ a small noise term. Such initial fluctuations (inevitable due to the intrinsic noise) can be amplified by reactions and propagated by diffusion. Under some conditions (Turing 1952; Murray, 1989) they can generate large-scale spatial structures with a characteristic scale. Such scale (wavelength) only depends on the intrinsic parameters involved in the RD terms. Such process of amplification of fluctuations can eventually shift the spatial distribution of morphogens from homogeneous to heterogeneous. Technically, this corresponds to a so called *symmetry breaking* (SB) phenomenon: the initial symmetry defined by the nearly homogeneous spatial state is broken. The conse-quence of such SB is a heterogeneous pattern of morphogen concentrations (Nicolis and Prigogine, 1977; Nicolis, 1995).

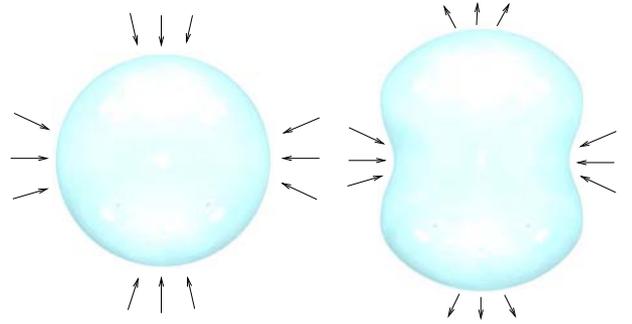

FIG. 1 Osmotic pressures in an ideal vesicle. These heterogeneous pressures can deform the membrane, eventually triggering membrane fission. Arrows indicate if the total pressure is compressive (the equator) or expansive (the poles). Models of cell replication must somehow create such spatially uneven pressure distribution.

Although the mechanisms underlying pattern formation in multicellular systems are typically richer than the previous RD scheme, they provide the appropriate framework to explain different situations. Some examples are pattern formation in fish (Kondo and Asai, 1995), bacterial growth in two dimensional cell cultures (Golding et al., 1998), sea shell patterns (Meinhardt, 1998), the skin of vertebrates (Suzuki et al., 2003; Maini 2003), the self-organization of ant cemeteries (Theraulaz et al., 2003) and the spatial distribution of population densities in ecosystems (Solé and Bascompte, 2006).

One particular scenario where living systems develop a spatial asymmetry is provided by single cells in morphogenesis. During early morphogenesis, cells often display a spatially asymmetric distribution of some molecules which appear preferentially located in different cell poles (Alberts et al., 2002). Such changes involve complex networks of molecular interactions and the reorganization of the cytoskeleton and are typically triggered by fertilization. At a simpler level, dynamical instabilities generating waves have been found in the cell cycle division of some bacteria. This seems to be the case of *Escherichia coli*, where a wave of protein concentrations, with rapid oscillations between the two membrane poles, seems to organize the division process (Raskin and de Boer, 1999; Hale et al., 2001). Although the full mechanism is rather complex and involves polymerization processes beneath the cytoplasmic membrane, the mechanism driving the cycle is simple.

The problem of how supramolecular assemblies self-reproduce is at the heart of current efforts directed towards building synthetic protocells (Szostack et al., 2001; Rasmussen et al., 2005). Although many works have been devoted to studying the chemical coupling between vesicles, enzymes and information molecules, they have so far failed to produce reliable self-replicating protocells, although significant steps have been performed (Luisi et al., 1999; Oberholzer and Luisi 2002; Nomura el al., 2003; Takakura el al., 2003, Hanczyc and Szostak, 2004; Noireaux and Libchaber, 2004; Deamer, 2005: Noireaux et al., 2005). One of the most promising approaches involves a top-down approximation using microscopic lipid vesicles incorporating preexisting biological molecules (Noireaux and Libchaber, 2004). The goal of this approach is finding a coupled set of reactions linking enzymes and/or information molecules with a container in such a way that growth and eventually reproduction can be achieved.

One problem with this top-down approach is that reliable processes leading to cell division need the appropriate coupling between the molecules involved (Solé et al., 2006). Such coupling must be able to increase cell size until some instability triggers vesicle splitting by generating a spatial breaking of symmetry through some active growth and deformation process. Properly designed, such process can eventually end up in vesicle splitting.

Here we propose a well-defined mesoscopic physico-chemical scenario of symmetry breaking instabilities which is shown to generate the appropriate, self-maintained spatial

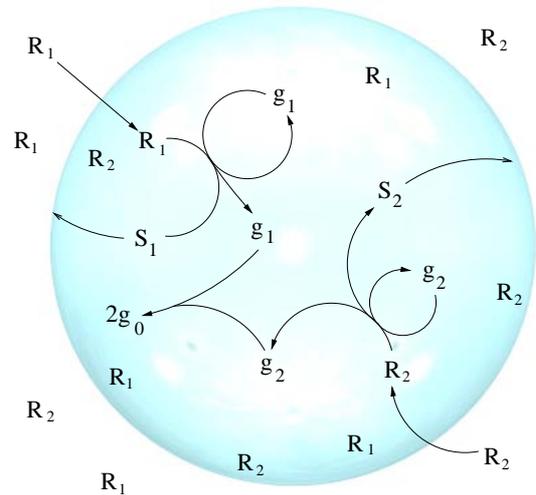

FIG. 2 The basic protocell model considered in this paper. It involves the presence of a membrane $\gamma$ together with precursors $(R_1, R_2)$ and two basic types of molecules (the morphogens $g_1$ and $g_2$) which interact and diffuse through the (two-dimensional) space, defining a minimal, spatially-extended metabolic network.

heterogeneities.

## II. PROTOCELL MODEL

The analysis of minimal cellular structures can contribute to a better understanding of possible prebiotic scenarios in which cellular life could have originated (Maynard Smith and Szathmáry, 2001) as well as to the design and synthesis of artificial protocells (Rasmussen et al., 2004). Considering a minimal cell structure, basically two mechanisms can give origin to reproduction: spontaneous division and induced division. Spontaneous division takes place when the vesicle grows until splitting into two daughter cells becomes energetically favorable. Induced division is a more complex mechanism that allows to internally controlling the division process. These two scenarios are very different. In most models of protocell replication, reactions are well described but the container appears only implicitly defined (Ganti, 2000; Kaneko and Yomo 2002; Munteanu and Solé, 2006; Kaneko, 2005, 2006; Sato and Kaneko, 2006) and thus an important ingredient of protocell dynamics (the explicit presence of a changing container) is missing. So far, models dealing with some type of self-reproducing spatial structure have been limited to special types lattice systems (Ono and Ikegami, 1999, Madina et al., 2003).

In an early work, the Russian biomathematician Nicolas Rashevsky (Rashevsky, 1960; Solé et al., 2006) explored the conditions for instability-induced cell division. He concluded that during the process of membrane growth there is a critical radius beyond which spontaneous division is energetically favorable. Moreover, he suggested that time and space-variable osmotic pressures were one of the most suitable mechanisms inducing membrane division.

In a previous work we have shown that osmotic-induced division is a feasible mechanism of vesicle self-replication (Macía and Solé, 2006). In this framework, the non-uniform distribution of osmotic pressures along the membrane is related to the non-uniform, enzyme-driven metabolite distribution inside the vesicle, with metabolic reactions taking place in specific locations, where metabolic centers are located. These centers (using the term coined by Rashevsky) could be specifically designed, synthetic trans-membrane proteins. This method, however, can trigger just a single vesicle division cycle. After division, only one metabolic center is present at each daughter cell and the division process cannot start again.

In our model (see below) replications take place indefinitely (provided that the appropriate precursors are available) and no enzymatic centers are required. A first approximation to such a minimal cell structure considers a continuous closed membrane involving some simple, internal metabolism. Here we propose a chemical mechanism coupled with vesicle growth, which generates the appropriate osmotic pressure distribution along the membrane (see figure 1). These pressure changes are generated by the interaction of Turing-like instabilities with vesicle dynamics. They are able to induce the correct vesicle deformation and eventually cell division.

Although previous work on pattern formation has already considered changing spatial domains due to tissue or organism growth (Meinhardt, 1982, Painter et al., 1992, Varea et al., 1997) and even membrane-bound Turing patterns (Levine and Rappel, 2005) as far as we know this is the first model where the boundaries are themselves a function of the reaction-diffusion dynamics, including membrane permeability and osmotic pressures altogether. As such, our model actually defines a totaly new class of spatially-extended dynamical system.

## III. METABOLISM-MEMBRANE SYSTEM

The problem of vesicle growth and division is not primarily thermodynamic, but kinetic (Morowitz, 1992). If a reliable cycle of cell growth and splitting is to be sustained, we need: (a) precursors provided from the external environment and (b) a restricted microenvironment where an appropriate set of reactions can drive the system out from equilibrium. Low-energy molecules and membrane precursors would be selectively transported across the membrane and high-energy compounds would be produced through energy-conversion processes.

Here we explicitly define all the components of our protocell model. The main goal is to introduce a set of reactions to be represented by a set of $n$ RD equations, namely:

$$\frac{dC_i}{dt} = \Phi_i(C_1, ..., C_n) + D_i \nabla^2 C_i \qquad (4)$$

with $i = 1, ..., n$ the index associated to the i-th morphogen. However, we will extend the formalism by incorporating a changing boundary which now acts as a permeable membrane, also coupled to the reactions described by $\Phi_i$. These reactions will define the protocell metabolism. Since osmotic pressures are associated to differences in molecular concentrations, active mechanisms generating spatial heterogeneity are expected to create changing pressure fields. These instabilities can break the osmotic pressure symmetry along the membrane, and after division the reactions defining the metabolism must be able to trigger a new growth-division cycle.

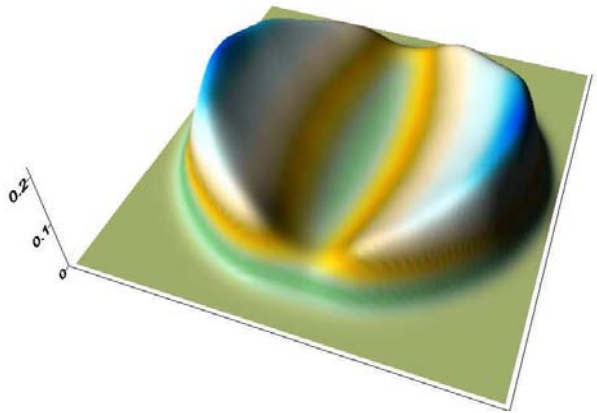

FIG. 3 Spatial distributions of morphogen concentrations $g_1$ and $g_2$, confined within a rigid circular container. Here the sum $g_1 + g_2$ is plotted, whit each morphogen concentrated in one of cell's poles. Non-uniform concentrations emerge from the autocatalytic effects of reactions (4) and (5) coupled with the inhibitory effects associated to reaction (6). Numerical calculations have been performed from a discrete integration of equations (7-11) and using the parameters in table I.

Let us first present the specific set of chemical reactions defining our basic metabolism. Several choices are possible and here we use one of the simplest scenarios found. As discussed by Morowitz, the logic of replication could be separated from chemical constraints, but in order to be able to test the feasibily of a given minimal protocell model, we should consider chemically reasonable sets of reactions to be implemented (Morowitz, 1992). Here we present such a simple, but chemically reasonable scenario. The set of reactions used here are:

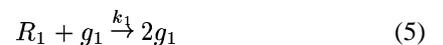
$$R_1 + g_1 \xrightarrow{k_1} 2g_1 \qquad (5)$$

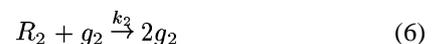
$$R_2 + g_2 \xrightarrow{k_2} 2g_2 \qquad (6)$$

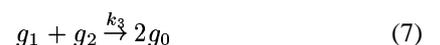
$$g_1 + g_2 \xrightarrow{k_3} 2g_0 \qquad (7)$$

In our model, reactions (4) and (5) can take place only inside the vesicle (due for example to the existence of inhibitory conditions outside it). Here $R_1$ and $R_2$ are the basic reagents, which are continuously and uniformly pumped from a source located at the limits of the system. The different substances involved in these reactions can cross the membrane with certain permeability and diffuse. The concentrations of the different molecules are denoted $c_z(r,t)$, where $z \in \{R_1, R_2, g_1, g_2, g_0\}$. For notational simplicity the spatio-temporal dependence is not explicitly written. The following set of partial differential equations describe the dynamics of the proposed system:

$$\frac{\partial c_{g_1}}{\partial t} = k_1 \cdot c_{R_1} \cdot c_{g_1} - \rho_1 \cdot c_{g_1} - k_3 \cdot c_{g_1} \cdot c_{g_2} + D_{g_1} \cdot \nabla^2 c_{g_1} \tag{8}$$

$$\frac{\partial c_{g_2}}{\partial t} = k_2 \cdot c_{R_2} \cdot c_{g_2} - \rho_2 \cdot c_{g_2} - k_3 \cdot c_{g_1} \cdot c_{g_2} + D_{g_2} \cdot \nabla^2 c_{g_2} \tag{9}$$

$$\frac{\partial c_{g_0}}{\partial t} = 2 \cdot k_3 \cdot c_{g_1} \cdot c_{g_2} - \rho_3 \cdot c_{g_0} + D_{g_0} \nabla^2 c_{g_0} \tag{10}$$

$$\frac{\partial c_{R_1}}{\partial t} = R_{1_0} - k_1 \cdot c_{R_1} \cdot c_{g_1} + D_{R_1} \nabla^2 c_{R_1} \tag{11}$$

$$\frac{\partial c_{R_2}}{\partial t} = R_{2_0} - k_2 \cdot c_{R_2} \cdot c_{g_2} + D_{R_2} \nabla^2 c_{R_2} \tag{12}$$

Here $\rho_1$, $\rho_2$ and $\rho_3$ are the degradation rates of $g_1$, $g_2$ and $g_0$ respectively. The diffusion coefficients are denoted by $D_{g_1}$, $D_{g_2}$, $D_{g_0}$, $D_{R_1}$ and $D_{R_2}$. For simplicity we assume that the value of the diffusion coefficients are the same inside and outside the membrane. Finally $R_{1_0}$ and $R_{2_0}$ are the constant rates of reagent supply.

This set of chemical reactions are able to trigger the emergence of a non-uniform spatial concentration of morphogens as a consequence of Turing-like instabilities. These instabilities are generated by the autocatalytic reactions (4) and (5) associated to the inhibitor effect of reaction (4). The previous model (7-11)) can be numerically solved by using a spatial discretization of the surface domain $\Gamma$, and considering zero-flux boundary conditions at the limits $\gamma$ of the domain, i. e. at the membrane.

We start with an initially homogeneous state where $C_z(r,t) = C_z^0 + \xi(r,0)$ with $C_z^0$ a constant and $\xi(r,0)$ a small noise term (Meinhardt, 1982). As a result of the previous set of interactions, concentrations change until they achieve a steady state. In figure 3 an example of the spatial distribution of $g_1$ and $g_2$ is shown, using the parameters given in table I. As expected from a symmetry-breaking phenomenon, the two morphogens get distributed in separated (exclusive) spatial domains. Each one tends to concentrate in one of the poles. These effects, coupled with membrane growth, will be exploited to design an active mechanism for controlled membrane division.

A second component of our model involves membrane growth. The cell membrane will grow as a consequence of the continuous input of molecules or aggregates available from an external source. As a consequence of this process, the boundary $\gamma$ (which now allows diffusion with the external enviroment) in not rigid anymore. At each time, $\gamma$ will be a time-dependent funciont $\gamma(t)$. The concentration at each instant depends on the number of molecules $n_j$ and the volume $V$. This dependence can be indicated as follows:

$$\frac{dc_j}{dt} = \frac{\partial c_j}{\partial n_j} \cdot \frac{\partial n_j}{\partial t}\bigg|_V + \frac{\partial c_j}{\partial V} \cdot \frac{\partial V}{\partial t}\bigg|_n \tag{13}$$

Considering that $c_j(t) = n_j(t)/V(t)$, equation (12) becomes:

$$\frac{dc_j}{dt} = \frac{\partial c_j}{\partial t}\bigg|_V - \frac{c_j}{V} \cdot \frac{\partial V}{\partial t}\bigg|_n \tag{14}$$

The first term in the right-hand side accounts for the change in concentrations associated to changes in the number of moles $n_j$ inside the protocell, assuming constant volume. The second term accounts for the change in concentration due to changes in membrane volume associated to membrane growth. If the composition of the external solution does not change over time, the rate at which the externally provided compounds are incorporated into the membrane can be considered proportional to its area $A$:

$$\frac{dA}{dt} = \frac{\ln 2}{T_d} A \tag{15}$$

where $T_d$ is the time required until membrane size is duplicated.

Cell volume changes due to both net water flow as well as to the growth of the membrane:

$$\frac{dV}{dt} = J_w A \tag{16}$$

In order to compute the net water in- and outflow, we must consider all the flows crossing the membrane. These flows are described by an additional set of equations. These equations account for the different interactions between all the elements of our system: water, solutes, and membrane (Kedem el al.

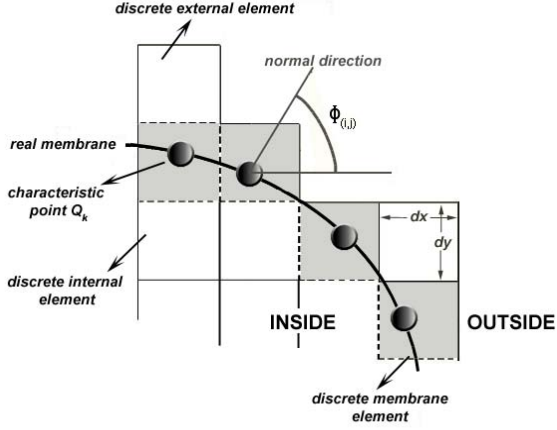

FIG. 4 Space discretization for the lattice model. The grid is formed by squares with a unit surface $\Delta S = \Delta x \Delta y$. There are tree types of squares or discrete elements: internal elements, covering the internal space surrounded by the membrane, external elements for the area outside the membrane and membrane discrete element.

1958, Patlak et al. 1963, Curry 1984). The net water flows can be expressed as follows:

$$J_w = L_p \left( \Delta p - RT \left( \sum_{j=1}^{2} \sigma_j \Delta c_j \right) \right) \tag{17}$$

where $L_p$ is the hydraulic conductivity of the water, $\Delta p$ is the hydrostatic pressure difference between the interior and the exterior, $R$ is the ideal gas constant, $T$ is the temperature, and $\sigma_j$ is the solute reflection coefficient for the $j$-th substance (zero for a freely permeable solute, and one for a completely impermeable one).

Moreover, to properly compute the concentration changes through time in equation (13) we must also take into account the total substance flow inwards or outwards the membrane. The flow can be expressed as (Curry 1984):

$$J_j = h_j \Delta c_j \left( \frac{P_e^j}{e^{P_e^i} - 1} \right) + J_w \left( 1 - \sigma_j \right) c_j^i \tag{18}$$

Here, $h_j$ is the permeability of the $j$-th substance (defined as the rate at which molecules cross the membrane), and

$$P_e^j = \frac{J_w \left( 1 - \sigma_j \right)}{h_j} \tag{19}$$

is the so called Peclet number.

## IV. MEMBRANE SHAPE

Membrane growth, as described by equations (14) and (15) will be affected by the Turing instabilities generated by reactions (4-6). The membrane expansion is followed by a loss of spatial symmetry due to the effects of non-uniform osmotic pressures along the membrane surface $\gamma(t)$.

The osmotic pressure at each membrane point is related to the current gradient of concentration between both membrane sides. At each point $r \in \gamma(t)$ the osmotic pressure value $P_j^o$ generated by the $j$-th substance can be written as:

$$P_j^o(r, t) = k_j (c_j^i(r, t) - c_j^e(r, t)) \tag{20}$$

where $k_j$ is a constant. For very low concentrations, we have $k_j = RT$, where $R$ is the ideal gas constant and $T$ is the temperature (if the concentrations are expressed in mols/liter). The osmotic pressure at one point $r$ of the membrane and at time $t$ can be calculated by adding the pressure generated by each substance, as follows:

$$P_t^o(r, t) = \sum_j k_j (c_j^i(r, t) - c_j^e(r, t)) \tag{21}$$

Finally we must take into account the contribution of the surface tension and the bending elasticity to the total pressure. This gives (for our two-dimensional system):

$$P_{tot}(r, t) = P_t^o(r, t) + \frac{2\gamma_0}{R_S(r)} + \frac{\kappa}{[R_S(r)]^2} \left( \frac{1}{R_S(r)} - \frac{1}{r_o} \right) \tag{22}$$

where $\gamma_0$ is the surface tension coefficient, and $k$ is the elastic bending coefficient. Here $R_s$(r) is the local radius of curvature, and $r_o$ is the spontaneous radius of curvature.

For simplicity we focus our attention on a two-dimensional model. The method employed to study the evolution of membrane shape has been previously presented in (Macía and Solé, 2006). This method considers only the local effect of osmotic pressures computed at each membrane point. Calculations are performed on a grid (Schaff et al., 2001). Figure 4 indicates how this discrete approximation is possible. In the lattice there are tree types of discrete elements: internal elements, which cover all the internal medium, external elements and membrane elements, which cover the real membrane and are in contact with both internal and external elements.

Membrane shape can be described by a set of *characteristic points* $Q_k$, each one associated to one membrane element. The position of each of these points can change dynamically as a consequence of pressure changes. In a first approximation, the displacement is proportional to the total pressure described by (21):

$$Q_k(t + \Delta t) = Q_k(t) + b \cdot P_{total}(t, Q_k) \tag{23}$$

Here $\Delta t$ is the discrete time interval used in the computation and $b$ is a constant. This constant value cannot be arbitrary. The position of the characteristic points defines the membrane shape and size. Such shape and size must be in agreement with the membrane size and volume as determined by equations (14) and (15).

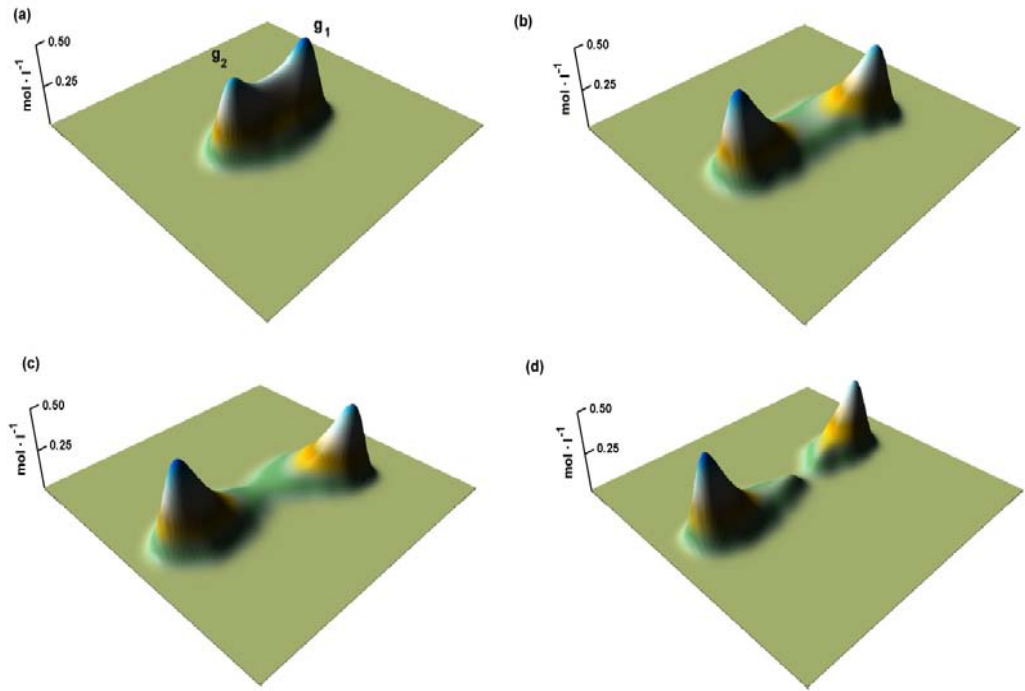

FIG. 5 Evolution of the concentrations profiles of $g_1$ and $g_2$ coupled with the membrane expansion process. Simulation parameters from table I. Here we can see that after a transient, two peaks emerge (a) indicating two maximal concentrations of $g_1$ and $g_2$. As the simulation proceeds, the peaks separate (b-c) as the membrane (not shown) gets deformed. In (d) we shaw the two concentration profiles right after cell splitting.

## V. RESULTS

The previous set of equations and boundary conditions allows the development of membrane growth and instability, using a realistic set of parameters (see table I). As discussed in section III, the system of chemical reactions generates steady non-uniform spatial concentrations. The question was to see if such spatial instabilities could trigger membrane changes leading to self-replication. In this section we summarize our basic results, showing that our model indeed displays the expected Turing-induced replication cycle.

As shown in figure 5, a spatial instability rapidly develops (figure 5a). The container where the metabolic reactions (4-6) are confined is a vesicle membrane. This membrane can be deformed and grow due the incorporation of new, externally provided, precursor molecules. The growth in membrane size, given by equation (14) in related with the volume increase given by equation (15). When cell area increases, the internal volume increases too, and the Turing instabilities move the maximal concentrations of $g_1$ and $g_2$ in opposite directions. Assuming that water and the different substances can cross the membrane with certain permeability, the non-uniform concentration distribution generates a non-uniform osmotic pressure along the membrane. Due this pressure, the membrane can be deformed with the characteristic shape described in figure 1. In this context, the metabolic reactions (4-6) have an active control on the membrane growth and shape.

Figure 5 shows the evolution of the concentration profiles of $g_1$ and $g_2$ coupled with the membrane expansion process.

As cell volume increases (in our 2D model this is represented by the internal area $|\Gamma|$) it enhances the spatial segregation of morphogens, due the increase in the size of the domain. In fact, in these regions where the concentrations of $g_1$ and $g_2$ are maximal there is a maximal expansive osmotic pressure. This pressure enhances the expansion in the poles, and as a consequence, the compression in the equatorial zone. The expansion taking place in the two poles enhances the separation of $g_1$ and $g_2$, and so on. The coupling of these two mechanisms, each one enhancing the other, creates a controlled membrane deformation. Beyond certain critical point membrane splitting becomes energetically favourable. In figure 6 we show the pressure distribution along the membrane in different simulation steps, consistently matching the theoretical pattern indicated in figure 1. Eventually, the narrow membrane division is a singularity and must be specifically introduced in the simulations (see Macía and Solé, 2006).

After division takes place, concentration distributions in each daughter cell are not anymore as those at the starting division cycle. Figure 5a shows the initial steps of the first cycle division, where the maximal concentration of $g_1$ and $g_2$ are comparable. However, figure 5d shows the situation right after division. In each cell there is a clearly dominant substance. Figure 7 shows the evolution on one daughter cell (the left one of figure 5d). As shown in figure 7(a-c) the minority substance regenerates and a new division cycle takes place. This is possible due the fact that, in spite of the kinetically symmetric features of both $g_1$ and $g_2$ (see Table 1), their respective reagents are different from the reagent of reaction (5).

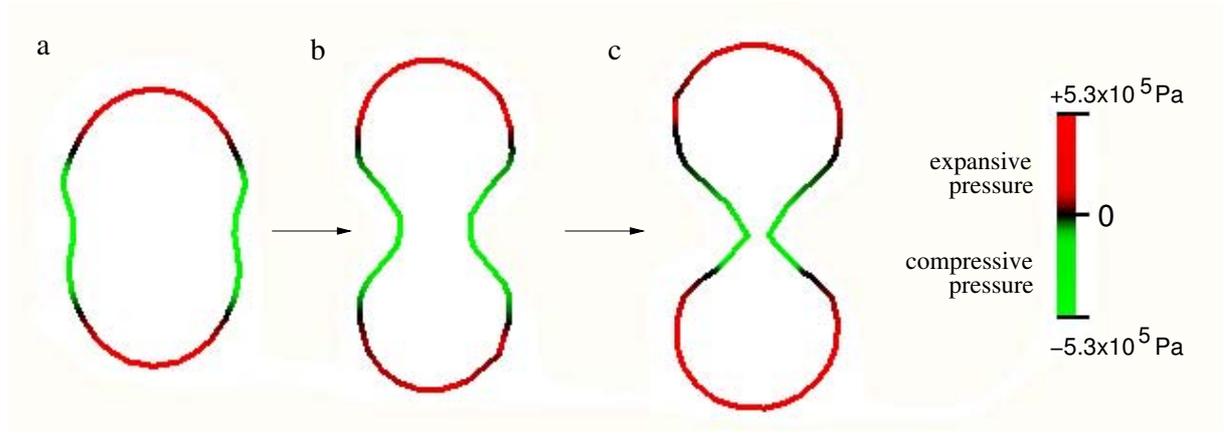

FIG. 6 Pressure distribution along the membrane $\gamma(t)$ at different simulation steps (resulting from the concentrations profiles of $g_1$ and $g_2$). Simulation parameters from Table I. Starting from a symmetric membrane and homogeneous concentrations of morphogens, the membrane gets elongated after a short transient (a) and starts to deform (b-c).

The concentration of the dominant morphogen ($g_2$ in figure 7a) has a growth limitation due to the substrate ($R_2$) depletion. This depletion is a consequence of the flow penetration limitation imposed by the membrane permeability. On the other hand, for the minority substance ($g_1$), the substrate consumption ($R_1$) is lower due the low concentration of the autocatalytic substance $g_1$. At the beginning, the inflow rate of $R_1$ is greater than its consumption. If the inhibitory effect of reaction (6) is not enough to limit the autocatalytic growth of $g_1$, the concentration can grow until the consumption of substrate $R_1$ is faster than the inflow rate. Then the depletion of substrate $R_1$ becomes the dominant mechanism and the growth of $g_1$ is limited. To accomplish these effects it is required that the diffusion coefficients of $R_1$ and $R_2$ are larger than the diffusion coefficient of $g_1$ and $g_2$.

These are common characteristics of Turing pattern formation in a wide range of scenarios (Meinhardt, 1982 and Murray, 1981) Furthermore, the permeability of $R_1$ and $R_2$ must be bigger than the permeability of $g_1$ and $g_2$ in order to accomplish a significant effect of the osmotic pressure in the poles. Finally, the kinetic constants $k_1$, $k_2$ must be greater than $k_3$ to ensure that the inhibitory effect of reaction (6) enhances the substances separation without preventing the growth of the minority substance (see Table I).

Our model assumes that the membrane remains continuously closed through time, as it grows following equation (14)[1]. In the model, the position of the characteristic points $Q_k$ determines the size and shape of the membrane. In order to have a physically consistent simulation, the size of the membrane calculated with (14) must be in agreement with the size as determined by the spatial location of the characteristic points. Figure 7(d-e) shows the dynamics of cell growth and replication in terms of membrane size. In figure 7(e) we can see the agreement between membrane size as calculated by (14) and the one derived from the characteristic point locations, as determined by (22). In figure 7(d) we display the membrane size evolution along three division cycles. The small differences between the consecutive cycles are an artifact of the model discretization.

TABLE I Parameters used in the simulations.

| Parameter | Symbol | Value |
|---|---|---|
| Kinetic constant | $k_1$ | $1.9 \times M^{-1} s^{-1}$ |
| " | $k_2$ | $1.9 \times M^{-1} s^{-1}$ |
| " | $k_3$ | $0.5 \times M^{-1} s^{-1}$ |
| Permeability | $h_{R_1}$ | $5 \times 10^{-6} cm s^{-1}$ |
| " | $h_{R_2}$ | $5 \times 10^{-6} cm s^{-1}$ |
| " | $h_{G_1}$ | $2 \times 10^{-9} cm \times s^{-1}$ |
| " | $h_{G_2}$ | $2 \times 10^{-9} cm \times s^{-1}$ |
| " | $h_{G_0}$ | $10^{-4} cm \times s^{-1}$ |
| Hydraulic conductivity | $L_p$ | $4.1 \times 10^{-13} cm Pa^{-1} s^{-1}$ |
| Diffusion coefficient | $D_{R_1}$ | $6 \times 10^{-8} cm^2 s^{-1}$ |
| " | $D_{R_2}$ | $6 \times 10^{-8} cm^2 s^{-1}$ |
| " | $D_{G_1}$ | $10^{-9} cm^2 s^{-1}$ |
| " | $D_{G_2}$ | $10^{-9} cm^2 s^{-1}$ |
| " | $D_{G_0}$ | $6 \times 10^{-8} cm^2 s^{-1}$ |
| Displacement proportionality constant | $b$ | $1.57 \times 10^{-13} cm Pa^{-1}$ |
| Substance supply | $R_{1_0}$ | $0.2 M$ |
| " | $R_{2_0}$ | $0.2 M$ |
| Surface Tension Coefficient | $\gamma_0$ | $1.98 Pa cm$ |
| Elastic bending coefficient | $k$ | $1.34 \times 10^{-19} Pa cm^3$ |
| Spontaneous radius of curvature | $r_0$ | $6 \mu m$ |
| Temperature | $T$ | $273 K$ |
| Degradation rate of $G_1$ | $\rho_1$ | $0.1 s^{-1}$ |
| Degradation rate of $G_2$ | $\rho_2$ | $0.1 s^{-1}$ |
| Degradation rate of $G_0$ | $\rho_3$ | $0.1 s^{-1}$ |

---

[1] Other parameters can be unable to keep the membrane growing or instead make it grow too fast. In those cases, membrane breaking can occur.

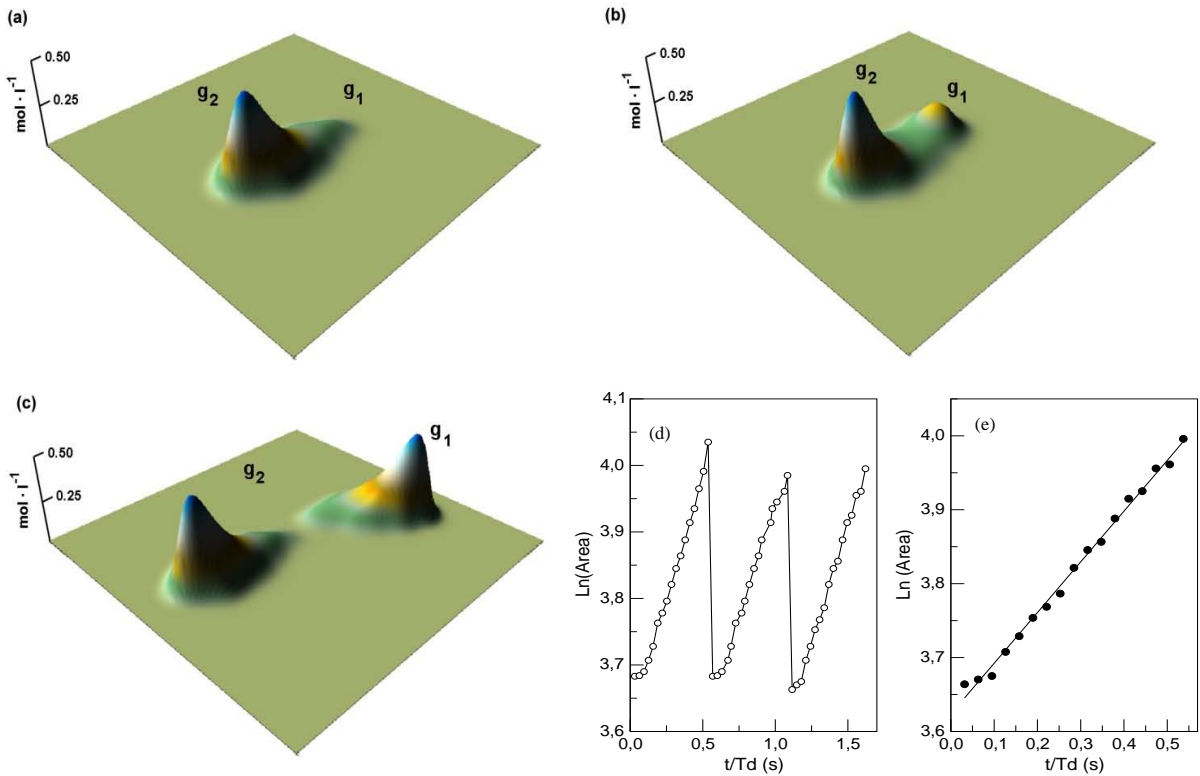

FIG. 7  (a-c) Evolution of the left daughter cell (starting from the final state shown in figure 5). The minority substance $g_1$ regenerates and a new division cycle takes place. In (d-e) we show the dynamics of cell growth and replication in term of membrane size. In (e) we can see the agreement between membrane size as calculated by (14) and one derived from the characteristic point locations, as determined by (22). In (d) we display the membrane size evolution along three division cycles. After each division only one daughter cell is represented. Here $T_d$ is the time necessary to duplicate the membrane size. Simulation parameters from Table I

## VI. DISCUSSION

We have presented a mesoscopic, minimal cellular model defined in terms of a closed membrane with a simple internal metabolism. These metabolic reactions are able to create Turing-like instabilities. Specifically, a mechanism leading to lateral inhibition associated to exclusive states has been used (Meinhardt, 1982). These instabilities, coupled with membrane growth, provide an active method for controlled membrane deformation and have been shown to trigger cell replication. The basic mechanism is related with a non-uniform osmotic pressure distribution along the membrane. Although a specific mechanism has been presented here, we have found other possible (more sophisticated) scenarios where this also occurs (Macia and Solé, unpublished).

Since spatial instabilities play an important role in many natural processes, the design of mechanisms based in these instabilities could be relevant to the synthesis of artificial protocells and even for understanding prebiotic scenarios of cellular evolution, were the sophisticated division mechanisms of the current cells were not present. In this context, the set of metabolic reactions can be arbitrarily generalized, thus opening the door for many different types of membrane-metabolism couplings. Future work should explore the possible paths towards the experimental synthesis of these protocells from available molecular structures, their potential evolvability and further extensions to more complex metabolic networks (Kaneko and Yomo, 2002).

## VII. ACKNOWLEDGMENTS

The authors thank the members of the Complex System Lab for useful discussions. Special thanks to Carlos Rodriguez-Caso for helpful comments and suggestions. This work has been supported by EU PACE grant within the 6th Framework Program under contract FP6-0022035 (Programmable Artificial Cell Evolution), by McyT grant FIS2004-05422 and by the Santa Fe Institute.